\documentclass[10pt,letterpaper]{article}
\usepackage{opex3}
\usepackage{epstopdf}
\usepackage{cite}
\begin{document}
\vskip4pc

\title{Nonlinear effects in multi-photon polaritonics}

\author{A. A. Pervishko,$^{1,*}$ T. C. H. Liew,$^{1,2}$ V. M. Kovalev,$^{3}$ I. G. Savenko,$^{1,4}$ and I. A. Shelykh$^{1,4}$},
\address{$^1$Division of Physics and Applied Physics, Nanyang Technological University, Singapore, 637371\\
$^2$Mediterranean Institute of Fundamental Physics, Rome, Italy, 00040\\
$^3$Russian Academy of Science, Institute of Semiconductor Physics, Siberian Branch, Novosibirsk, Russia, 630090\\
$^4$Science Institute, University of Iceland, Reykjavik, Iceland, IS-107}
\email{$^*$anas0005@e.ntu.edu.sg} 

\begin{abstract} We consider theoretically nonlinear effects in a semiconductor quantum well embedded inside a photonic microcavity.
Two-photon absorption by a 2p exciton state is considered and investigated; the matrix element of two-photon absorption is calculated.
We compute the emission spectrum of the sample and demonstrate that under coherent pumping the nonlinearity of the two photon absorption process gives rise to bistability.
\end{abstract}


\section{Introduction}
Polaritonics is a rapidly developing branch of science lying at the intersection of the physics of semiconductors, quantum mechanics and nonlinear optics. The interest in this field has been stimulated by a series of fascinating experimental and theoretical discoveries, which demonstrate that a variety of quantum collective phenomena can be observed in polariton systems including: Bose-Einstein condensation (BEC)~\cite{NatureKasprzak,Balili,Lai} and superfluidity~\cite{AmoSuperfluidity,AmoSuperfluidity2,CarusottoSuperfluidity}; the Josephson effect~\cite{Josephson}; and the formation of various topological excitations such as vortices~\cite{Vortex,Lagoudakis,Nardin} and solitons~\cite{AmoSolitons,Grosso,Sich,HalfSoliton}, as well as other spatial patterns~\cite{Manni,Christmann,Kammann}.
While most fundamental studies have been performed at low temperature, experiments in GaN~\cite{Christopoulos,Ayan}, ZnO~\cite{ZnO} and organic~\cite{Kena-Cohen} microcavities demonstrate devices working at room temperature. Now it becomes obvious, that polaritonics has a huge potential for technological applications~\cite{PolDevices}.

Polaritonics stems from the proposal that quantum microcavities can be used as efficient sources of THz radiation~\cite{Kavokin1}, which needs efficient pumping of the 2p excitonic state. It can not be excited under single photon absorption due to optical selection rules. In one of the recent publications~\cite{KavokinPRL}, some theoretical aspects of THz radiation emission under two-photon excitation were considered. The consideration was based on the master equation approach. The two-photon absorption and THz emission were investigated as incoherent processes. This kind of consideration, however, is not always applicable. One can imagine the situation when the two-photon transition is tuned in resonance with the energy of the confined electromagnetic mode of a cavity. In this case, the coupling is of resonant character and the theoretical description should be performed along another route. Theoretical consideration of this regime is the subject of the present paper.

We consider a cavity which is subject to two-photon resonant excitation, in which takes place, pumping of the dark (2p) exciton state. Instead of considering the strong coupling between near resonant photons and 1s excitons~\cite{Weisbuch}, we uncover a different kind of exciton-polariton generated by the non-linear coupling of photon pairs with 2p excitons. Similiar to the conventional case, a characteristic splitting of the photoluminescence spectrum is observed. However, the value of the Rabi splitting now becomes dependent on the intensity of the pump due to the nonlinear nature of the two photon absorption. This nonlinearity also results in the onset of bistable behavior. 

\section{Polaritonics with 2p dark exciton state}

We consider the following geometry (Fig. 1): a quantum well (QW) with 2p excitonic transition having energy $\hbar\omega_p$ placed inside a planar microcavity supporting the photonic mode of energy $\omega_{2c}\approx\omega_p/2$. In this configuration the cavity mode can resonantly interact with the 2p excitonic state. 

\begin{figure}
\centering
    \includegraphics{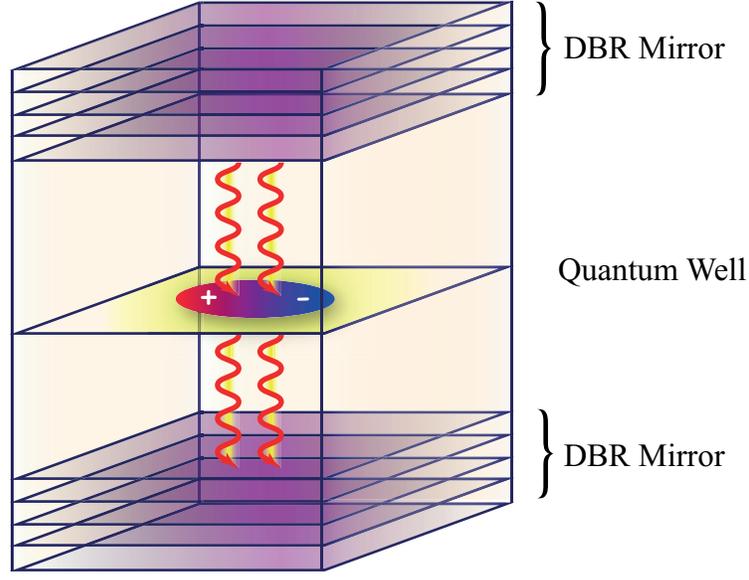}
\caption{Geometry of the structure. We consider a microcavity, which is made by Bragg mirrors, with the quantum well inside, where a $2p$ excitonic state can be excited by two photons(red curves) with energy $\hbar\omega_{2c}$ each.}
\end{figure}

The Hamiltonian of the system reads:

\begin{equation}
\hat {\cal H}=\hbar\omega_p \hat p^\dagger \hat p+\hbar\omega_{2c}\hat a^\dagger \hat a+g\left(\hat p\hat a^\dagger \hat a^\dagger+\hat p^\dagger \hat a\hat a\right),
\end{equation}

where the operators $\hat p,\hat a$ correspond to 2p dark excitons and photons respectively and satisfy bosonic commutation relations. The first two terms in Eq. (1) describe free excitons and photons and the last term corresponds to the resonant interaction between them, where $g$ is the two photon-exciton coupling constant
calculated in Appendix using second-order perturbation theory:

\begin{eqnarray}
g=\sqrt{\frac{S}{2}}{\left(-\frac{qA_0}{\mu}\right)}^2 \sum_{n}\frac{\frac{i\sqrt{E_g m_0^2}}{\sqrt{2m^*}}\Phi_n(0)\int d\vec{r}R_{21}(\vec r)\vec r R_{10}(\vec{r})}{2\omega - (E_g -E_n+\omega)}\frac{im_0}{\hbar\sqrt{2}}(E_{2p}-E_{ns}),
\end{eqnarray}

where $S$ is a quantization area of the sample; $m_0$, $m^*$ and $\mu$ are the free electron, effective and reduced exciton masses, respectively; $q$ is the elementary charge of the electron; $E_g$ is the band gap of the material; $\omega$ is the energy of the photon; $E_n$ are the eigenvalues of the 2D Hydrogen atom; $\Phi_n(\vec r)$ and $R_{21}(\vec r)$, $R_{10}(\vec r)$ are normalized angular and radial eigenfunctions of the 2D Hydrogen atom, which are written explicitly in the Appendix, Eqs. (30) and (31), respectively, and $\vec{A}$ is the vector potential, which can be writen in the dipole approximation:

\begin{eqnarray}
\vec A=\vec e A_0,~~~~~A_0=\sqrt{\frac{\hbar}{2\epsilon\epsilon_0\omega L S}},
\end{eqnarray}

where $\epsilon_0$ and $\epsilon$ are vacuum and material permitivities; $L$ is the length of the cavity. 

\subsection{Energy levels for the case of an ideal cavity}

Note, that the Hamiltonian (Eq. (1)) commutes with the excitations' number operator $\hat{N}=\hat a^\dagger \hat a+2\hat p^\dagger \hat p$, that means that the eigenvectors of these two operators coincide and the Hilbert space of the problem can be represented as a direct product of the manifolds corresponding to given numbers of $N$. Each manifold consists of $\lfloor N/2 \rfloor$ basis vectors: $|N,0\rangle,|N-2,1\rangle,....,|0,N/2\rangle$ for even $N$ and $|N,0\rangle,|N-2,1\rangle,....,|1,N-1/2\rangle$ for odd $N$. Here the first number in kets corresponds to the number of photons, and the second number to the number of 2p excitons. The matrix element of Hamiltonian (Eq. (1)) between any two vectors belonging to manifolds with different $N$ is zero. This means that determination of the spectrum of the problem consists in diagonalization of $N\times N$ matrices, each of which has a finite size.

For $N=0$ and $N=1$ the result is trivial and the corresponding energies are $E_0=0$ and $E_1=\hbar\omega_{2c}$ respectively. The cases of $N=2,3$ are also straightforward to consider, as the size of the reduced Hamiltonian for the manifold is 2$\times$2 in these cases:
\begin{eqnarray}
{\cal{H}}_2=\left(\begin{array}{cc}
  2\hbar\omega_{2c} & g\sqrt{2} \\
  g\sqrt{2} & \hbar\omega_p
\end{array}\right), ~~~~~{\cal{H}}_3=\left(\begin{array}{cc}
  3\hbar\omega_{2c} & g\sqrt{6} \\
  g\sqrt{6} & \hbar(\omega_p+\omega_{2c})
\end{array}\right),
\end{eqnarray}
and the eigenenergies in the case of resonance $2\omega_{2c}=\omega_p$ are
\begin{equation}
E_2=2\hbar\omega_c\pm g\sqrt{2}, ~~E_3=3\hbar\omega_c\pm g\sqrt{6}.\nonumber
\end{equation}
Consideration of the states with larger numbers of photons requires diagonalization of matrices of higher size. The energy spectrum is represented in Fig.~2(a).

\begin{figure}
    \includegraphics{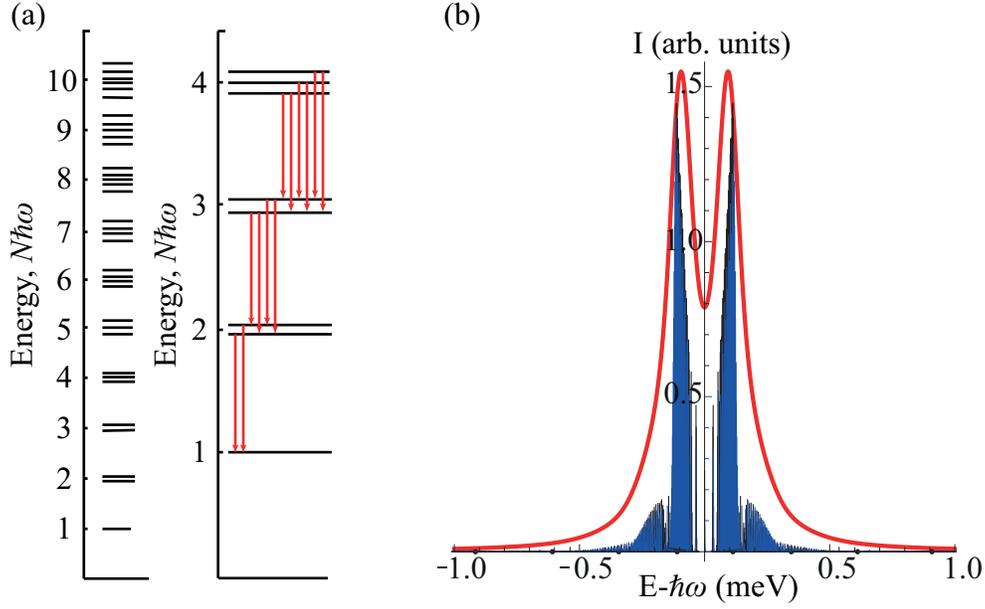}
\caption{(a) Energy levels of the QW with 2p excitonic transition placed inside the microcavity. We consider the case when the exciton energy is $\hbar\omega_{p}$ and the photon energy is $\omega_{2c}\approx\omega_p/2$. Possible transitions for $N=1,2,3,4$ manifolds for the ideal cavity are represented on the right side of the figure (a). (b) Emission spectrum of the system. The intensity of transitions from individual levels is presented by the blue lines. We consider the case when $\omega_{2c}\approx\omega_p/2=0.71 eV, g=0.025 meV$ and the average number of the excitations in the system $\langle{N}\rangle=30$ and the statistics of photons is Poissonian. Including a Lorentzian broadening of each transition line, gives the double peak spectrum of emission shown by the red curve. We consider broadening equals $0.05meV$. One can define the value of the Rabi splitting in the system as the distance between the two peaks.}
\end{figure}

\subsection{Spectrum of emission}
 If the mirrors of the cavity are not perfect and some photons tunnel through them, the system can emit in the outside world. The spectrum of emission is determined by the energy distances between the levels corresponding to $N$ and $N-1$ manifolds and the intensities of the transitions will be proportional to squares of the matrix elements of the photon annihilation operator $I_{if}\sim|\langle\Psi^{out}|\hat a|\Psi^ {in}\rangle |^2\color{black}p(\Psi^ {in})$, 
where $|\Psi^ {in}\rangle$ is an initial wavefunction in the manifold with N cavity photons and $|\Psi^{out}\rangle$ is a final wavefunction in the manifold with (N-1) cavity photons. $p(\Psi^{in})$ represents the probability of occupation of the initial state \cite{Kibis2012}. Likewise assume that the number of excitations in the system is sufficiently to apply the central limit theorem. In this case the occupation of states with different numbers of particles is described by a Poisson distribution:
\begin{equation}
p(\lambda; N)=\frac{\lambda^N e^{-\lambda}}{N!}.
\end{equation}
Figure~2(b) shows the total photoluminescence emission spectrum.

It is well-known, that in the case of one photon and a resonant 1s bright exciton in a standard semiconductor microcavity the energy splitting does not depend on the occupation number and equals $\Delta\Omega_{R,S}=2g_s$, if $g_s$ is the bright exciton- single photon coupling strength. Considering our system with two-photon absorption for large $N$ ($N>>1$), and substituting $a,a^\dagger\rightarrow\sqrt{N}$ one immediately obtains $\Delta\Omega_R\approx 2g\sqrt{N}$. Accounting for finite lifetime of the photons and excitons, one obtains~\cite{Microcavity}:

\begin{equation}
\Delta\Omega_R\approx\sqrt{4g^2N-\frac{(\gamma_a-\gamma_p)^2}{4}}.
\end{equation}

Varying the number of the excitations, which can be done by increasing of the pump power, one can thus change the Rabi splitting in the system. Note that coupling constant g scales with the quantization area as $g\sim(S)^{-1/2}$, so $g^2N$ depends on the concentration of the photons $n_a=N/S$ only.

\section{Classical field equations for coherent pump}
To investigate the dynamics of the system under coherent pump with account for finite lifetimes of the photons and dark excitons in the cavity, the approach should be modified. We will use a system of coupled nonlinear equations for the dynamics of the macroscopic wavefunctions describing cavity photons and dark excitons. The latter can be obtained in the following way. First, let us write the Heisenberg equations of motion for the operators $\hat p$ and $\hat a$:

 \begin{eqnarray}
 i\hbar\frac{d\hat p}{dt}&=&\left[\hat p,\hat {\cal{H}}\right]=\hbar\omega_p\hat p+g\hat a^2,\\
 i\hbar\frac{d\hat a}{dt}&=&\left[\hat a,\hat {\cal{H}}\right]=\hbar\omega_a\hat a+2g\hat p\hat a^\dagger.
\label{motion2}
 \end{eqnarray}

Then, one should take averages of the above equations thus passing from the operators to the mean values as $\hat p\rightarrow\langle \hat p\rangle=Tr\{\rho \hat p\}$. We present a mean product of several operators by products of the mean values of the operators (e.g. $\langle \hat p \hat a^ \dagger\rangle\rightarrow\langle \hat p\rangle\langle \hat a\rangle^\ast, \langle \hat a^2\rangle\rightarrow\langle \hat a\rangle^2$). This approximation is relevant for coherent statistics.
Finally, we phenomenologically introduce the lifetimes of the modes and external coherent pumping of cavity photons (by a coherent laser beam) with frequency $\omega$. As a result,  Eqs.~(8) and (9) transform into the following system of nonlinear differential equations:

\begin{eqnarray}
\frac{d\hat p}{dt}&=&-(i\omega_p+\frac{\gamma_p}{\hbar})\hat p-i\Omega_p\hat a^2,\\
\frac{d\hat a}{dt}&=&-(i\omega_a+\frac{\gamma_a}{\hbar})\hat a-2i\Omega_p\hat p\hat a^\ast-\frac{i}{\hbar}Pe^{-i\omega t}.
 \end{eqnarray}
 
 where $\gamma_{p,a}=1/(2\tau_{p,a})$ with $\tau_{p,a}$ being lifetimes of the modes, $P$ is the amplitude of the coherent pump and $\Omega_p=g/\hbar$. Making a substitution $\hat a\leftarrow \hat ae^{-i\omega t}, \hat p\leftarrow \hat pe^{-2i\omega t}$, in the stationary regime the equations read:
 \begin{eqnarray}
&&(i\hbar\Delta_p+\gamma_p) p+i\hbar\Omega_p a^2=0,\\
&&(i\hbar\Delta_a+\gamma_a) a+2i\hbar\Omega_p p a^\ast=-iP,
\end{eqnarray}
 where $\Delta_p=\omega_p-2\omega,\Delta_a=\omega_a-\omega$, which can be reduced to:
 \begin{eqnarray}
(i\hbar\Delta_a+\gamma_a) a+\frac{2\hbar^2\Omega_p^2}{i\hbar\Delta_p+\gamma_p}|a|^2 a+iP=0.
 \end{eqnarray}
 This result can also be written in terms of the real functions describing the number of the photons, $N_a=|a|^2$ and excitons, $N_p=|p|^2$:
 \begin{eqnarray}
&&N_a\left[1+c_1N_a+c_2N_a^2\right]=I_a,\\
&&N_p=\frac{g^2N_a^2}{\gamma_p^2+\hbar^2\Delta_p^2},
 \end{eqnarray}
 where
 \begin{eqnarray}
&&c_1=\frac{4g^2(\gamma_a\gamma_p-\hbar^2\Delta_a\Delta_p)}{(\hbar^2\Delta_p^2+\gamma_p^2)(\hbar^2\Delta_a^2+\gamma_a^2)},\\
&&c_2=\frac{4g^4}{(\hbar^2\Delta_p^2+\gamma_p^2)(\hbar^2\Delta_a^2+\gamma_a^2)},\\
&&I_a=\frac{|P|^2}{\hbar^2\Delta_a^2+\gamma_a^2}.
 \end{eqnarray}

Equation~(15) is a cubic equation for $N_a$. The results are represented in Fig.~3(a) for the following parameters corresponding to the system: $\gamma_a=0.05$ $meV$, $\gamma_p=0.01$ $meV$,  and three different values of detunings $\Delta_a$ and $\Delta_p$. These curves reveal an S-shaped dependence of the mode populations on the intensity of the pump, which characterizes the phenomenon of bistability, which is well-known in semiconductor microcavities~\cite{Kulakovski2000, Gippius2004, Whittaker2005, Baas2004, Gippius2007, Paraiso2010, Bajoni2008}. If the value of the pumping intensity lies in the bistable zone, the system can, in principle, occupy more than one possible state, and the particular choice of this state depends on the history of the evolution.

Bistability is a fundamental ingredient of several devices based on semiconductor microcavities, such as memory elements~\cite{Shelykh2008}, spin switches~\cite{Amoswitch}, spin patterns~\cite{Sarkar,Adrados}, and optical circuit designs~\cite{Liew2008,Liew2010}.
\begin{figure}
\centering
\includegraphics[width=370pt, height=145pt]{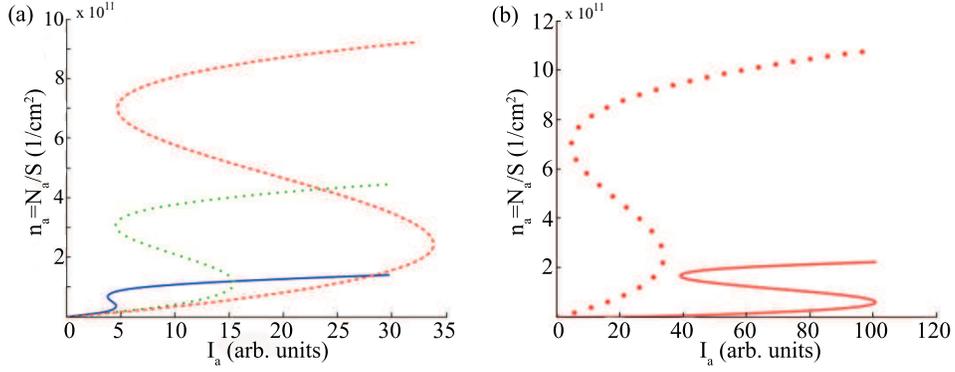}
\caption{Considering the finite lifetime of the photons and external coherent pumping of the cavity, we observed a bistable behaviour of the number of photons on intensity of the coherent pump. (a) The parameters are varied for detuning of the system: (red dashed line) $\hbar\Delta_a=0.375meV, \hbar\Delta_p=0.75  meV$; (green dot line) $\hbar\Delta_a=0.25meV, \hbar\Delta_p=0.5 meV$; (blue bold line) $\hbar\Delta_a=0.125 meV,\hbar \Delta_p=0.25meV$.
The influence of the exciton-exciton interaction on the hysteresis curve is illustrated in figure (b), where the cases of $\alpha_p$=0 (red dashed curve) and $\alpha_p=6E_Ba^2_B/S$ (red solid curve) are presented. The detuning parameters are $\hbar\Delta_a=0.375meV, \hbar\Delta_p=0.75  meV$.}
\centering
\end{figure}
However, there is substantial difference between the previous works and the paradigm developed in our paper. Indeed, in most of the previous approaches bistability resulted from the exciton-exciton (Coulomb and exchange) interaction, even in THz emitting systems~\cite{Savenko2011}.
Here, the bistability arises due to the nonlinearity of the two-photon absorption process. It should be noted that, still, excitons are present in the system and their interaction can, in principle, change the physical behavior. The exciton-exciton interaction can be accounted for by means of the term:
  \begin{equation}
  {\cal{H}}_{int}=\frac{\alpha_p}{2} \hat p^\dagger \hat p^\dagger \hat p\hat p, \nonumber
  \end{equation}
 which should be added to Hamiltonian (Eq. (1)).
Then, the equations of motion change to:
\begin{eqnarray}
\frac{d\hat p}{dt}=-(i\omega_p+\frac{\gamma_p}{\hbar})\hat p-i\Omega_p\hat a^2 -\frac{i}{\hbar}\alpha_p \hat p^{\dagger} \hat p\hat p,\\
\frac{d\hat a}{dt}=-(i\omega_a+\frac{\gamma_a}{\hbar})\hat a-2i\Omega_p\hat p\hat a^\ast-\frac{i}{\hbar}Pe^{-i\omega t},
 \end{eqnarray}
and in the stationary regime we obtain the following solution:
\begin{eqnarray}
n_a+\frac{4((\gamma_a \gamma_p-\hbar^2\Delta_a\Delta_p)n_p-\hbar\Delta_a\alpha_p n_p^2)}{(\hbar^2\Delta_a^2+\gamma_a^2)}+\frac{4g^2n_pn_a}{(\hbar^2\Delta_a^2+\gamma_a^2)}=I_a.
\end{eqnarray}

To define the constant $\alpha_p$, we use the formula $\alpha_p\approx6E_Ba_B^2/S$, where $E_B$ is the exciton binding energy, $a_B$ is the exciton Bohr radius and $S$ is the laser spot area~\cite{Tassone}. In QWs based on GaAs-based alloys, we can use the parameters: $E_B\approx19.2$ $meV$ and $a_B=5.8$ $nm$. The result is presented in Fig. 3(b).

As one can see the exciton-exciton interaction term changes the shape of the bistable behavior of the system. It becomes more extended on intensity of the coherent pump. The density of the excitons compared with the non-interacting case decreases as well. We therefore conclude that the influence of the nonlinearity of the two-photon absorption on the bistability of the system is comparable with the contribution of exciton-exciton interactions. For simplicity, here we have considered a single spin degree of freedom, which corresponds to the case in which a circularly polarized optical pump excites 2p excitons with a single spin polarization. Indeed, we do expect a variety of spin sensitive phenomena to take place in the system with different polarizations of pumping. Note that the polarization of 2p excitons can still be controlled via the polarization of the optical pumping, allowing analogous effects to those in conventional microcavities~\cite{Ivan}.

\section{Conclusions}
 We developed a theoretical approach for the description of strong light-matter interaction in a system with two-photon absorption processes. We calculated the photoluminescence spectrum, where a splitting occurs due to the coupling between photons and 2p dark-excitons in the two photon absorption process. We developed the mean-field equations, valid under coherent excitation of the system, and demonstrated that the nonlinearity of the two-photon absorption process gives rise to the phenomenon of bistability. The two-photon to 2p exciton matrix element was accurately calculated. This formalism is not only directly applicable for device construction, but also highlights the nonlinear nature of the system, which may be useful for the realization of quantum optical devices under coherent pumping~\cite{Verger, Jacobs}.

\section{Appendix}
In the present appendix we calculate the matrix element of two-photon absorption using second order perturbation theory. The excitation of the 2p exciton state can occur, mediated by a virtual state representing the state of the system after the absorption of a photon. The virtual states of s-type are the only ones allowed by selection rules and the transition from an s-type state to the 2p exciton state occurs with the absorption of the second photon. The total two-photon absorption matrix element is given by summing over all virtual s-type states. 

The Hamiltonian in the case of exciton-photon interaction can be written in the form:
\begin{equation}
\hat {\cal{H}}=\frac{(\hat{\vec{p_e}}-q_e\vec{A_e})^2}{2m_e}+\frac{(\hat{\vec{p_h}}-q_h\vec{A_h})^2}{2m_h}+U(\vec{r_e}-\vec{r_h}),
\end{equation}
where $q_e$ and $q_h$ are the elementary charges of an electron and hole, respectively; $m_e$ and $m_h$ are the electron and hole effective masses, respectively; $U(\vec{r_e}-\vec{r_h})$ is the electron-hole interaction potential; $\hat{\vec{p_e}}$ and $\hat{\vec{p_h}}$ are the quasi-momenta; $\vec{A}$ is the vector potential of the electromagnetic field.
In the dipole approximation, the vector potential does not depend on coordinate and reads:
\begin{eqnarray}
\vec{A_e}=\vec{A_h}=\vec{e}A_0,~~~~~A_0=\sqrt{\frac{\hbar}{2\epsilon\epsilon_0\omega V}},
\end{eqnarray}
where $\vec{e}$ is a unit vector describing the light polarization. Here $\epsilon_0$ and $\epsilon$ are the vacuum and the relative permittivity, respectively; $\omega$  is the frequency of exciting light; $V$ is the cavity volume, $V\approx (\lambda/2n)S$, where  $\lambda$ is the wavelength of the fundamental cavity mode; $n$ is the refractive index of the material and $S$ is the area of the laser spot which roughly defines the area in which excitons exist. Keeping the terms linear in $\vec A$ and introducing the center of mass coordinates, we obtain the equation for the perturbation operator:
\begin{equation}
\hat W=-\frac{qA_0}{\mu}\vec{e}\cdot\hat{\vec{p}},
\end{equation}
where $\hat{\vec{p}}$ is the quasi-momentum of the relative electron-hole motion, $q=|q_e|=|q_h|$ and $\mu=m_e m_h/(m_e+m_h)$ is the reduced mass. The intrinsic motion of the exciton can be described by the Hamiltonian:
\begin{equation}
\hat {\cal{H}}=\frac{\hat{\vec p}^2}{2\mu}+U(r).
\end{equation}
We will not consider the motion of the center of mass, as in dipole approximation it does not feel any influence of light at normal incidence. The eigenfunctions and eigenstates will be similar in form to eigenfunctions and eigenstates for the 2D Hydrogen atom with the difference that the reduced mass '$\mu$' replaces the free electron mass, the material permitivity replaces the vacuum permitivity and energies are taken with respect to the band-gap~\cite{AnalytHydrogen}. The eigenstates for a 2D systen are described by the Schr\"{o}dinger equation, which in polar coordinates reads:
\begin{eqnarray}
\left( -\frac{\hbar^2}{2\mu}\left[\frac{\partial^2}{\partial r^2}+\frac{1}{r}\frac{\partial}{\partial r}+\frac{1}{r^2}\frac{\partial^2}{\partial \phi^2}\right]-\frac{q^2}{4\pi\epsilon_0\epsilon r} \right)\psi(r,\phi)=(E-E_g)\psi(r,\phi),
\end{eqnarray}
where $E_g$ is the semiconductor band-gap.

Using separation of variables
\begin{equation}
\psi(r,\phi)=R(r)\Phi(\phi),
\end{equation}
we obtain the normalized angular and radial eigenfunctions:
\begin{eqnarray}
\Phi(\phi)=\frac{1}{(2\pi)^{1/2}}e^{il\phi},\quad  l=0,\pm 1,\pm 2,...\\
R_{nl}(r)=\frac{\beta_n}{(2|l|)!}\left[\frac{(n-|l|-1)!}{(2n-1)(n-|l|-1)!}\right]^{1/2}(\beta_n r)^{|l|}e^{-\beta_n r/2} {_1F_1}(-n+|l|+1,2|l|+1,\beta_n r),
\end{eqnarray}
where $n$ and $l$ are principle and angular momentum quantum numbers, respectively;

\begin{equation}
\beta_n = \frac{2}{4\pi\epsilon_0\epsilon(n-1/2)}\frac{\mu e^2}{\hbar^2},
\end{equation}
and $_1F_1$ is the confluent hypergeometric function. 

The 2D eigenvalues are:
\begin{equation}
E_n=E_g-\frac{1}{2(4\pi\epsilon_0\epsilon)^2 (n-1/2)^2}\frac{\mu q^4}{\hbar^2}.
\end{equation}

The process we are interested in goes through the virtual state, and the transition can be described by means of second-order perturbation theory. The corresponding matrix element is:
\begin{equation}
M_{fi}=\sum_{\xi}\frac{\langle f|\hat W|\xi\rangle\langle \xi|\hat V^{(1)}|i\rangle}{E_i-E_{\eta}}+{\langle f|\hat V^{(2)}|i \rangle}.
\end{equation}
Here $\hat V^{(1)}$ and $\hat V^{(2)}$ are perturbation operators of the first and second order, respectively.  In Eq. (34) summation over $\xi$ means summation over $\eta$ (the virtual levels) and integration over momentum. Here $i$ is the initial state: there are 0 excitons, 2 photons in the system. It can be described by:
\begin{eqnarray}
|i\rangle=\delta(\vec{r_e}-\vec{r_h}),~~~~~E_i=2\omega,
\end{eqnarray}
$\eta$ is a virtual state with one exciton and one photon:
\begin{eqnarray}
|\eta\rangle=F_{\eta}(\vec{r_e},\vec{r_h})=\frac{e^{i\frac{\vec{P_c}}{\hbar}\vec{R_c}}}{\sqrt{S}}\psi_{\eta}(\vec{r}),~~~~~ E_\eta=E_g-E_n+\omega.
\end{eqnarray}
Here $f$ denotes a final state, which describes excitation of the 2p level. Substituting the wavefunctions into the equation for the matrix elements, one can obtain:
\begin{eqnarray}
\langle \eta|\hat V^{(1)}|i \rangle&& =-\frac{qA_0}{\mu}(\vec{e}\cdot\hat{\vec{p}})_{cv}\int d\vec{r_e}\int d\vec{r_h}\delta(\vec{r_e}-\vec{r_h})F_{\eta}(\vec{r_e};\vec{r_h})=\nonumber\\
&&=-\frac{qA_0}{\mu}(\vec{e}\cdot\hat{\vec{p}})_{cv}R_{\eta}(0) \frac{1}{\sqrt{S}}(2\pi\hbar)^2\delta(\vec{P_c}),\\
\langle f|\hat W|\eta \rangle &&=-\frac{qA_0}{\mu}\int d\vec{R_c}\int d\vec{r}F^*_{\eta'}(\vec{r_e};\vec{r_h})\vec{e}\cdot\hat{\vec{p}}F_{\eta}((\vec{r_e};\vec{r_h}))=\nonumber\\
&&=-\frac{qA_0}{\mu}\delta(\vec{P'_c}-\vec{P_c})\frac{1}{S}(2\pi\hbar)^2\int d\vec{r}R^*_{\eta'}(\vec{r})\vec{e}\cdot\hat{\vec{p}}R_{\eta}(\vec{r})=\nonumber\\
&&=-\frac{qA_0}{\mu}(\vec{e}\cdot\hat{\vec{p}})_{\eta'\eta}\frac{1}{S}(2\pi\hbar)^2\delta(\vec{P'_c}-\vec{P_c}),
\end{eqnarray}
where $(\vec e\cdot\hat {\vec p})_{cv}$ is the matrix element of transition between valence and conduction band Bloch functions. The matrix element $\langle f|\hat V^{(2)}|i \rangle$ is equal to zero as a consequence of orthogonality of the Bloch amplitudes of the conduction and valence bands. 

The full matrix element takes the form:
\begin{equation}
M_{fi}=\sqrt{S}{\left(-\frac{qA_0}{\mu}\right)}^2 \sum_{\eta}\frac{(\vec{e}\hat{\vec{p}})_{\eta'\eta}(\vec{e}\hat{\vec{p}})_{cv}\psi_{\eta}(0)}{2\omega - (E_g -E_n+\omega)}.
\end{equation}
First, let us find the interband matrix element $\langle \eta|\hat V^{(1)}|i\rangle$.
Built on the Bloch functions, this matrix element can be estimated from the $\mathbf{k}\cdot\mathbf{p}$ method:
\begin{equation}
\frac{1}{m^*}\approx \frac{2|(\vec{e}\hat{ \vec{p}})_{cv}|^2}{E_gm_0^2},
\end{equation}
or using rudimentary considerations,
\begin{eqnarray}
i\hbar\frac{d \hat x}{d t}=i\hbar\frac{\hat p}{m_0}=\left[\hat x; \hat {\cal{H}}\right] ~~~\Rightarrow~~~
\langle c|\hat p|v\rangle\approx\frac{i}{\hbar} m_0aE_g.
\end{eqnarray}
Thus,
\begin{equation}
{(\vec{e}\hat{\vec{p}})}_{cv}\approx\frac{i}{\hbar}E_g a m_0,
\end{equation}
where $E_g$ is the bandgap and $a$ is the lattice constant.

Second, one should find the term $(\vec{e}\vec{p})_{\eta'\eta}$.
In polar coordinates, the momentum operator can be presented as:
\begin{equation}
\hat p=-i\hbar\hat\bigtriangledown=-i\hbar\left(\hat r \frac{\partial}{\partial r}+\hat {\phi}\frac{1}{r}\frac{\partial}{\partial \phi}\right),
\end{equation}
and for circularly polarized light:
\begin{equation}
\vec e \cdot\hat{\vec p}=-\frac{i \hbar}{\sqrt{2}}e^{i\phi}\left(\frac{\partial}{\partial r}+i\frac{1}{r}\frac{\partial}{\partial \phi}\right).
\end{equation}
Due to angular momentum conservation, the possible $\eta$ states have the orbital momentum equal to zero. The radial wavefunction is then:
\begin{eqnarray}
R_{n0}&&=\beta_n\frac{1}{\sqrt{2n-1}}e^{-\frac{\beta_n r}{2}}{_1F_1}(-n-1,1,\beta_n r)=\nonumber\\
&&=\beta_n\frac{1}{\sqrt{2n-1}}e^{-\frac{\beta_n r}{2}}L^0_{n-1}(\beta_n r).
\end{eqnarray}
We are interested in the case when the final state is the 2$p$ state, $n=2,l=1:$
\begin{equation}
R_{21}=\frac{\beta_2^2}{\sqrt{6}}re^{-\frac{\beta_2 r}{2}}.
\end{equation}
Using definition of momentum we can switch from the momentum to coordinate operator:
\begin{eqnarray}
\langle 2p|\hat p| 1s\rangle=\langle 2p|m_0\frac{i}{\hbar}[\hat {\cal{H}},\hat r]|1s\rangle=\frac{im_0}{\hbar}(E_{2p}-E_{1s})\langle 2p|\hat r|1s\rangle. 
\end{eqnarray}
As a result, the matrix element can be presented in form:
\begin{eqnarray}
\langle f|\hat W|\eta \rangle &&=-\frac{qA_0}{\mu}\delta(\vec{P'_c}-\vec{P_c})\frac{(2\pi\hbar)^2}{S}\delta(\vec{P'_c}-\vec{P_c})\int d\vec{r}R^*_{\eta'}(\vec{r})\vec{e}\hat{\vec{p}}R_{\eta}(\vec{r})=\nonumber\\
&&=-\frac{qA_0}{\mu}\delta(\vec{P'_c}-\vec{P_c})\frac{(2\pi\hbar)^2}{S}\int d\vec{r}R^*_{\eta'}(\vec{r})\frac{e^{i\phi}}{\sqrt{2}}\left(i\hbar\frac{\partial}{\partial r}\right )R_{\eta}(\vec{r})=\nonumber\\
&&=-\frac{qA_0}{\mu}\delta(\vec{P'_c}-\vec{P_c})\frac{(2\pi\hbar)^2}{S}\int d\vec{r}\frac{e^{i\phi}}{\sqrt{2}}\frac{im_0}{\hbar}\left(E_{\eta'}-E_{\eta}\right )R^*_{\eta'}(\vec{r})\vec r R_{\eta}(\vec{r}),\\
\langle 2p|\hat W|ns \rangle &&=-\frac{qA_0}{\mu}\delta(\vec{P'_c}-\vec{P_c})\frac{e^{i\phi}}{\sqrt{2}}\frac{im_0}{\hbar}\left(E_{2p}-E_{ns}\right)\frac{(2\pi\hbar)^2}{S}\int d\vec{r}R^*_{2p}(\vec{r})\vec r R_{1s}(\vec{r}).
\end{eqnarray}

The overlap integral can be calculated as:
\begin{eqnarray}
\int d\vec{r}R_{21}(\vec r)\vec r R_{n0}(\vec{r})=\frac{243(n-2)^{n-3}(n+1)^{-n-2}(2n-1)^5}{(\mu q^2/(\pi\epsilon_0\epsilon\hbar^2))^4}\frac{\beta^2_2\beta_n}{\sqrt{6(2n-1)}}.
\end{eqnarray}
We did not take into account $n=2$, since the energy splitting between the $2s$ and $2p$ levels is insignificant compared to the other levels (causing ($E_{2p}-E_{2s}$) to vanish). 
Finally,
\begin{eqnarray}
M_{2p,i}=\sqrt{S}{\left(-\frac{qA_0}{\mu}\right)}^2 \sum_{n}\frac{\frac{i\sqrt{E_g m_0^2}}{\sqrt{2m^*}}\Phi_n(0)\int d\vec{r}R_{21}(\vec r)\vec r R_{10}(\vec{r})}{2\omega - (E_g -E_n+\omega)}\frac{im_0}{\hbar\sqrt{2}}(E_{2p}-E_{ns}).
\end{eqnarray}
For a GaAs structure, $a=5.6$ $\AA$, $S= \pi R^2$, $R=0.1$ $\mu m$, $E_g=1.42$ $eV$, $\epsilon=12.9$, $n=3.312$, $\lambda=800$ $nm$, $\mu=0.058m_0$, $m^*=0.067m_0$. This matrix element is connected with the interaction constant '$g$', which describes interaction between photons and exciton in the Hamiltonian ${\cal{H}} = g\hat p\hat a^{+}\hat a^{+}+g\hat p^+\hat a\hat a$. 
\begin{equation}
M_{2p,i}=\langle 2p|\hat {\cal{H}}|2\omega \rangle=g\langle 2p|\hat p\hat a^+a^++\hat p^+\hat a\hat a|2\omega\rangle=g\sqrt{2}.
\end{equation}
It is approximately equal to $0.025$ meV.

\section*{Acknowledgments}
The work was supported by IRSES "POLATER" AND "POLAPHEN" projects. T. C. H. Liew was supported by the EU Marie-Curie project EPOQUES. V. M. Kovalev was supported by RFBR 12-02-31012. I. G. Savenko acknowledges support of the Eimskip foundation.

\end{document}